\documentclass[sigconf]{acmart}
\setcopyright{none}
\renewcommand\footnotetextcopyrightpermission[1]{}

\AtBeginDocument{\DeclareCaptionSubType{lstlisting}}

\usepackage[all]{nowidow}
\usepackage[english]{babel}
\usepackage[strict]{changepage}
\usepackage[T1]{fontenc}
\usepackage[utf8]{inputenc}
\usepackage[dvipsnames]{xcolor}
\usepackage{amsmath,amsfonts}
\usepackage{amsthm}
\usepackage{anyfontsize}
\usepackage{array}
\usepackage{booktabs}
\usepackage{color}
\usepackage{comment}
\usepackage{etoolbox}
\usepackage{float}
\usepackage{graphicx}
\usepackage{hyperref}
\usepackage{hyphenat}
\usepackage{lineno}
\usepackage{listings}
\usepackage{makecell}
\usepackage{mdframed}
\usepackage{multirow}
\usepackage{setspace}
\usepackage{stfloats}
\usepackage{subcaption}
\usepackage{syntax}
\usepackage{tcolorbox}
\usepackage{textcomp}
\usepackage{url}
\usepackage{verbatim}
\usepackage{xspace}
\usepackage{colortbl}
\usepackage{array}
\usepackage{multirow}
\usepackage{ragged2e}
\usepackage{colortbl}
\usepackage{hhline}
\usepackage[linesnumbered]{algorithm2e}
\usepackage{tabularx}

\hypersetup{
    colorlinks=true,
    linkcolor=blue,
    filecolor=magenta,
    urlcolor=black,
    citecolor=violet,
    pdfpagemode=FullScreen,
    }

\definecolor{light-gray}{gray}{0.85}

\lstdefinelanguage{diff}{
  morecomment=[f][\textit]{@@},
  morecomment=[f][\color{red}]{-},
  morecomment=[f][\color{green!60!black}]{+},
  morecomment=[f][\textit]{---},
  morecomment=[f][\textit]{+++},
}

\mdfsetup{
  skipabove=0.5em,
  skipbelow=0.5em,
  innertopmargin=0.5em,
  innerbottommargin=0.5em
}

\let\oldnl\nl
\newcommand{\nonl}{\renewcommand{\nl}{\let\nl\oldnl}}

\newcommand{\revision}[1]{{\color{black}#1\normalfont}}
\newcommand{\toolname}{\textsc{SafeRefactorPy}\xspace}

\newcommand{\totalTargets}{1,152}

\newcommand{\totalBugsBehavioral}{13}

\newcommand{\oss}{gpt-oss-20b}

\newcommand{\code}[1]{\texttt{#1}}

\usepackage{tikz}

\microtypesetup{activate={true,nocompatibility}, patch=none}

\acmConference[SBES 2026]{40th Brazilian Symposium on Software Engineering}{September 8–11, 2026}{São Paulo, SP, Brazil}

\begin{document}

\title{Detecting Behavioral Changes in Python Refactoring Implementations with Foundation Models}

\author{Jonhnanthan Oliveira}
\affiliation{%
  \institution{Federal University of Campina Grande}
  \city{Campina Grande, PB}
  \country{Brazil}
}
\email{jonhnanthan@copin.ufcg.edu.br}

\author{Rohit Gheyi}
\affiliation{%
  \institution{Federal University of Campina Grande}
  \city{Campina Grande, PB}
  \country{Brazil}
}
\email{rohit@dsc.ufcg.edu.br}

\author{Márcio Ribeiro}
\affiliation{%
  \institution{Federal University of Alagoas}
  \city{Maceió, AL}
  \country{Brazil}
}
\email{marcio@ic.ufal.br}

\author{Alessandro Garcia}
\affiliation{%
  \institution{Pontifical Catholic University of Rio de Janeiro}
  \city{Rio de Janeiro, RJ}
  \country{Brazil}
}
\email{afgarcia@inf.puc-rio.br}

\renewcommand{\shortauthors}{Oliveira et al.}
\renewcommand{\shorttitle}{Detecting BC in Python Refactoring Implementations with Foundation Models}

\begin{abstract}
Python is a widely adopted programming language, valued for its simplicity and flexibility.
However, automated refactoring for Python remains challenging, even though refactoring is an essential practice in software evolution aimed at improving internal code structure without changing external behavior.
Understanding how behavioral changes are introduced during refactoring is crucial, as such issues can compromise software reliability and reduce developer productivity.
We propose an approach based on a foundation model oracle that analyzes git-style diffs to identify behavioral changes introduced by Python refactorings.
We evaluated our technique on Rope refactoring implementations, reusing \totalTargets{} refactoring attempts from a prior study and analyzing 217 resulting transformation pairs with the oracle.
Our model-based analysis uncovered \totalBugsBehavioral{} distinct bugs among the seven refactoring types studied.
All reported bugs were submitted to the respective developers, and 12 of the 13 resulting issue reports were accepted according to issue-tracker evidence.
These results highlight the need to improve the robustness of current Python refactoring tools to ensure the correctness of automated code transformations and support reliable software maintenance.
\end{abstract}


\frenchspacing
\maketitle

\section{Introduction}

Python is a widely used programming language, valued for its simplicity and flexibility~\cite{Rossum-book-2011}.
Its syntax supports multiple paradigms, such as imperative, object-oriented, and functional, and allows developers to choose between typed and untyped code.
While these features contribute to Python's popularity, they also complicate code maintenance and tooling support, especially for automated refactoring~\cite{Fowler-book-1999,Opdyke-PHD-1992,Mens-TSE-2004}.

Refactoring is a key practice in software evolution that aims to improve internal code structure without changing external behavior.
Despite this simple definition, implementing correct and reliable refactoring transformations is non-trivial~\cite{Tempero-ACM-2017,Schafer-OOPSLA-2010,tip-oopsla-2003,schafer-ecoop-2009,schafer-oopsla-2008,Steimann-ecoop-2009}.
This task becomes particularly challenging in languages such as Python~\cite{Schafer-fwrt-2012}, where dynamic typing, runtime name resolution, reflection, and dynamic dispatch can make transformations difficult to validate statically and may lead to runtime errors or unintended behavior.
Consequently, a refactoring transformation may appear to succeed from the tool's perspective while still changing the observable behavior of the resulting program.

Prior research has introduced techniques for evaluating the correctness of refactoring tools, especially in statically typed languages.
For instance, Sch\"{a}fer et al.~\cite{Schafer-OOPSLA-2010} developed an intermediate representation to simplify the specification and verification of Java refactorings.
Mongiovi et al.~\cite{Mongiovi-icsme-2014,Mongiovi-TSE-2018} and Soares et al.~\cite{Soares-TSE-2013,Soares-ICSM-2011} investigated compilation errors, behavioral changes, and overly strong preconditions in Java-based refactoring engines.
In addition, AlOmar et al.~\cite{AlOmar-infsof-2021} mapped behavior-preservation techniques over the years, showing that much of the existing evidence has focused on Java and other statically typed settings.

For Python, only recent work has started to expose limitations in refactoring support.
Wang et al.~\cite{Wang-icsme-2025} report that Python refactoring tools still face practical limitations in coverage and robustness, reinforcing the need for better tooling support in this ecosystem.
Oliveira et al.~\cite{sbes2025} investigated faulty Python refactorings and identified problems related to type errors and other statically observable failures.
However, these studies do not address the complementary problem of detecting behavioral changes in apparently successful Python refactoring transformations.
This distinction is important because behavioral changes may not manifest as syntax, type, or name-resolution failures, and may only become visible when the transformed program is executed or inspected semantically.
There is still a need to understand how Python refactoring tools affect behavior.
In particular, existing studies do not provide a Python-specific evaluation of behavioral-change detection under a diff-based protocol on real-world refactoring instances, which limits direct evidence on whether apparently successful transformations preserve observable behavior in practice.

This problem is relevant because developers increasingly rely on Integrated Development Environments (IDEs) and automated tools to perform refactorings. There is also an increasing demand for improved tooling support among Python developers~\cite{golubev-fse-2021,Wang-icsme-2025}. Such tools must handle Python's dynamic features without introducing behavioral regressions. Even widely used refactoring engines can introduce subtle behavioral changes~\cite{sun-ase-2022,Mongiovi-TSE-2018,Soares-TSE-2013,Soares-ICSM-2011,Oliveira-2019-IST,Schafer-PLPV-2009,Pinto-wrt-2013,Tip-TOPLAS-2011,daniel-icse-2023}, which may go unnoticed by developers and result in runtime errors or degraded software quality~\cite{Cedrim-2017-FSE}. This highlights the need for validation mechanisms that can inspect apparently successful refactoring transformations and flag behavior-changing edits before they are integrated into development workflows.

In this paper, we \revision{extend} \toolname{}~\cite{sbes2025} \revision{with} a technique that uses a foundation model-based oracle to analyze git-style diffs and detect behavioral changes introduced by Python refactorings.
The key idea is to use the diff as a compact representation of the transformation: instead of requiring the full project as input, \toolname{} focuses the oracle on the concrete edits performed by the refactoring and asks whether those edits preserve observable behavior.
This design makes the approach lightweight and portable across refactoring tools, since integrating a new tool mainly requires changing the step that applies refactorings and generates diffs.
At the same time, \toolname{} is intended as a bug-detection and validation aid rather than a replacement for test suites, static analyses, or project-wide semantic checks.

\revision{We evaluated \toolname{} by reusing refactoring transformations produced with Rope~\cite{rope} in our prior work~\cite{sbes2025}.}
\revision{We analyzed 217 program pairs from that dataset to detect behavioral changes.}
In the main evaluation, conducted with \oss{}, we compared the oracle's classifications against a human-expert baseline using standard classification metrics and repeated-run stability measures.
The oracle achieved high recall, indicating that the approach is useful for exposing behavior-changing transformations, although false positives show that model-based judgments should complement rather than replace conventional validation techniques.
We also include an exploratory complementary analysis with GPT-5.2 Thinking to examine how a stronger proprietary model changes the reasoning scope of the oracle. 

This process identified \totalBugsBehavioral{} distinct, hard-to-identify bugs among the seven frequent refactoring types studied.
Although 217 transformation pairs do not support broad prevalence claims about all Python refactoring tools, our study is designed as a focused bug-finding and oracle-evaluation study.
Its main evidence comes from the distinct root causes uncovered, their diversity across refactoring types, and their external validation by developers.
We reported all discovered bugs to the relevant developers.
According to issue-tracker evidence based on comments, labels, and issue status, 12 of the 13 reports were accepted as bugs.
These results suggest that diff-based foundation-model oracles can provide practical value as an additional validation layer for Python refactoring tools, especially in scenarios where behavioral changes are subtle and undesirable while being difficult to encode as static rules.

\section{Motivating Example}
\label{sec:motivatingExample}

This section presents a motivating example of a behavioral change silently introduced by a refactoring in Python, illustrating why detecting such changes requires reasoning about observable behavior. 
Consider the Python program in Listing~\ref{lstMotivatingBC:input}, which defines a class \code{WordList} that extends the built-in \code{list} class.
The class overrides the \code{extend} function to ensure that every string element appended to the list is first converted into a \code{Word} instance through the custom \code{append} function.
When executing this program, the output confirms that all elements stored in the list are \code{Word} instances.

Suppose a developer decides to apply the Inline Method refactoring using Rope~\cite{rope} to the \code{extend} function of \code{WordList}, since its body simply delegates to \code{self.append} in a loop.
According to the refactoring specification~\cite{Fowler-book-1999}, Inline Method replaces calls to the function with its body and removes the original function declaration.
Rope applies the transformation \revision{and removes the original function declaration} without emitting any warning\revision{, as shown in lines 10--12 in Listing~\ref{lstMotivatingBC:input}}.

\begin{lstlisting}[language=diff, caption={A git-style diff of a Python program.}, label=lstMotivatingBC:input,
    numbers=left,
    numberstyle=\tiny,
    stepnumber=1,
    numbersep=6pt,
    xleftmargin=2.2em,
    framexleftmargin=2.2em,
    belowskip=0.1mm,
]
class Word(str):
    def __new__(cls, value):
        return super().__new__(cls, value)
class WordList(list):
    def append(self, obj):
        if isinstance(obj, str)
                and not isinstance(obj, Word):
            obj = Word(obj)
        super().append(obj)
-   def extend(self, iterable):
-       for e in iterable:
-           self.append(e)
wl = WordList()
wl.extend(["hello", "world"])
print([type(x).__name__ for x in wl])
\end{lstlisting}

The refactored program executes without any errors.
However, the observable behavior has changed: the elements stored in the list are now \code{str} instances instead of \code{Word} instances.
This occurs because, after the removal of the overriding \code{WordList.extend} function, the call \code{wl.extend(["hello", "world"])} now resolves to the inherited \code{list.extend()}, which adds elements directly without invoking the custom \code{append} function that performs the conversion to \code{Word}.
Both the original and the refactored programs are syntactically correct, and no runtime exception is raised.
Yet, the program produces different observable behavior, which may propagate as a subtle regression in downstream code that depends on the elements being \code{Word} instances.

This bug is documented in Rope's issue tracker.\footnote{\url{https://github.com/python-rope/rope/issues/828}}
It illustrates a fundamental limitation of checking only whether a transformation succeeds: behavioral changes can be introduced even when the resulting program remains executable.
In such cases, reasoning about whether the refactoring preserves semantics requires understanding the \emph{intent} of the code and the interactions between overridden functions and their callers.
Previous work has proposed techniques to detect behavioral changes in refactorings for other languages~\cite{Soares-TSE-2013,Mongiovi-TSE-2018,Soares-ICSM-2011}, but these techniques do not directly address Python-specific mechanisms such as dynamic dispatch and runtime method resolution. Thus, these techniques fall short of detecting undesirable behavioral changes in the presence of such mechanisms.
To address this gap, in Section~\ref{sec:technique}, we present a foundation model-based oracle designed to detect behavioral changes by analyzing git-style diffs of refactoring transformations.

\section{Detecting Behavioral Changes in Refactorings}
\label{sec:technique}

This section details how we extended \toolname{} with git-style diffs and a foundation-model oracle to detect behavioral changes caused by Python refactorings.

\subsection{Overview and Refactoring Application}
\label{subsec:techOverview}

\begin{figure*}[t]
    \centering
    \includegraphics[keepaspectratio, width=0.9\textwidth]{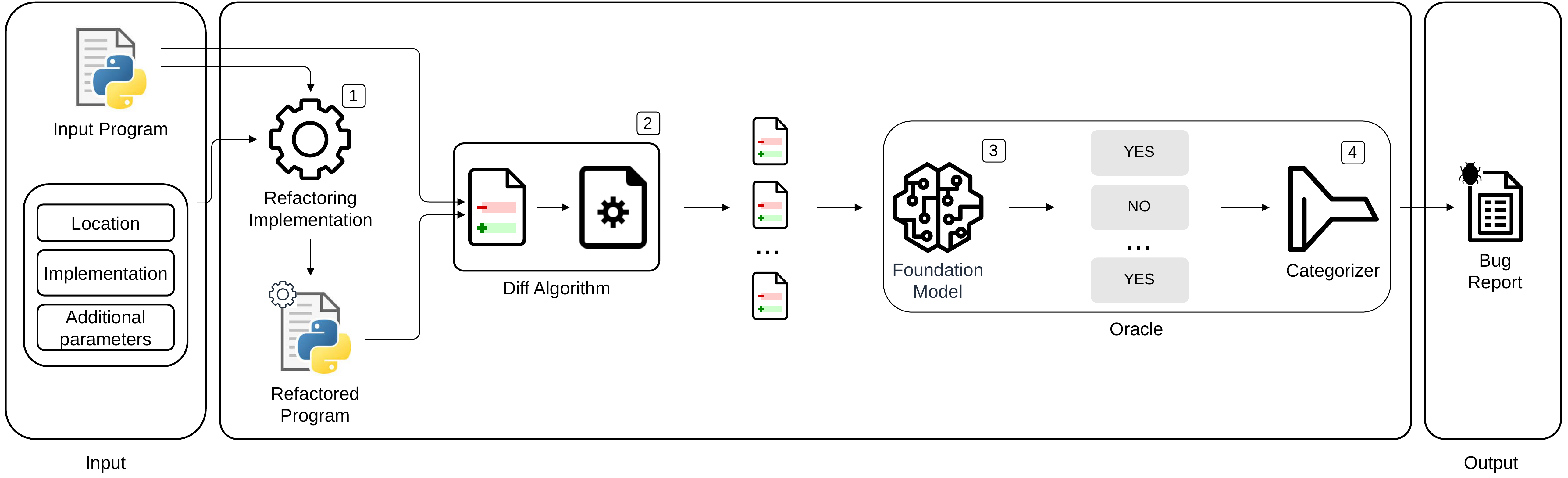}
    \caption{\toolname{}: a model-based pipeline for detecting behavioral changes introduced by Python refactorings.}
    \label{fig:technique}
    \Description{}
\end{figure*}

\paragraph{Overview.} Our technique takes as input a Python program, the refactoring implementation under test, the target location, and any required parameters (Step~1). If the transformation succeeds, it computes a git-style diff between the original and refactored programs (Step~2) and submits that diff to a foundation-model oracle that decides whether the instance preserves observable behavior (Step~3). After all instances have been analyzed, the categorization algorithm aggregates the results, identifies recurring failure patterns, and assembles bug reports for expert inspection (Step~4). If the refactoring tool throws an exception, this outcome is recorded directly in the bug report.
Figure~\ref{fig:technique} provides an overview of the behavioral-change detection pipeline.

\paragraph{Refactoring Application.} The technique applies the selected refactoring to the input program using the target location and any required parameters.
\revision{For example, in Inline Method (Section~\ref{sec:motivatingExample}), the inputs are the program and the function location.}
Other refactoring types follow the same pattern with adjusted parameters. This is the only tool-specific step of the pipeline: integrating a different IDE or refactoring engine requires changing only how refactoring instances are applied.

\subsection{Diff Generation}
\label{subsec:techStep3}

After applying the refactoring, our technique computes a git-style diff between the original and refactored versions of the program.
This diff is used as input for the behavioral-change detection phase.
\revision{The oracle receives only this diff; it does not receive the full source file or the complete project.}
Listing~\ref{lst:motivate-diff} shows a representative git-style diff.

We use git-style diffs as the oracle input for three reasons.
First, diffs provide a compact representation of the transformation, which helps address the context-window and inference-cost limitations of foundation models.
Feeding an entire project to the model may be impractical for real-world codebases, whereas the diff exposes the concrete edits introduced by the refactoring.
Second, diffs focus the oracle on the transformation itself, reducing irrelevant project context and making the analysis closer to how developers inspect refactoring changes in code review.
Third, the format is tool-independent: any refactoring engine that produces an original and a resulting program can be integrated by generating the corresponding diff.

This design is particularly useful for Python refactorings, where behavior-preservation violations may arise from small edits involving special methods, dynamic dispatch, name resolution, imports, or indentation-sensitive scopes.
By presenting the model with the precise set of modifications, \toolname{} prioritizes the detection of behavioral changes that may be missed by traditional precondition checks or by simply observing that the transformation completed successfully.
This choice is also a methodological trade-off: using diffs improves scalability, cost, and portability, but it may miss effects that depend on project context not visible in the diff, such as external callers, subclass interactions, or import-time dependencies.

\subsection{Detecting Behavioral Changes}
\label{subsec:techStep6}

In this step, we invoke a foundation-model oracle to determine whether the applied refactoring introduces behavioral changes. We consider behavior preservation in terms of observable behavior: a refactoring should preserve the program's outputs, exceptions, externally visible state, timing or synchronization behavior, and other externally observable effects.

We use zero-shot prompting~\cite{prompts,prompt-techniques,zero-shot-prompt}. The prompt was built from domain knowledge and prior refactoring literature~\cite{Opdyke-PHD-1992,Roberts-PHD-1999,Fowler-book-1999}.
We refined the initial prompt through an iterative meta-prompting process~\cite{reynolds2021prompt,hou-metaprompting,zhang2023meta}. In each iteration, we submitted the current version of the prompt, together with a set of representative diffs and their expected ground-truth labels, to an auxiliary foundation model acting as a \emph{prompt critic}.
The critic was instructed to identify ambiguities, missing edge cases, or unclear instructions that could lead to incorrect classifications. Based on this feedback, we revised and re-evaluated the prompt on the same representative diffs, for example by clarifying assumptions about consistent renames outside the diff, comment-only edits, and common behavior-changing categories.
We repeated this revise–evaluate cycle until the prompt achieved stable and correct classifications on all representative cases.
The final prompt is the result of this refinement process.

The model is instructed to act as a refactoring-aware code assistant and to judge behavioral preservation based only on a git-style diff. We define behavior preservation in terms of observable outputs, exceptions, visible state, and related external effects~\cite{Opdyke-PHD-1992,tsantalis-tse-2009,Roberts-PHD-1999}. The oracle, therefore, complements rather than replaces unit tests: it can flag risky transformations without executing the program.

We also instruct the model to treat code outside the diff as unchanged, avoid inventing symbols or files, and answer in a restricted plain-text format that our categorizer can parse reliably. Both the oracle and the human baseline are intentionally diff-scoped. The oracle is model-agnostic; any suitable foundation model can be plugged into this step.
In the prompt, \code{diff} represents the Git-style diff as shown in Listing~\ref{lst:motivate-diff}.

\begin{lstlisting}[
  language=diff,
  caption={A git-style diff example of the application of Rope Rename Method refactoring.},
  label={lst:motivate-diff}
]
--- a/blob.py
+++ b/blob.py
@@ -67,7 +67,7 @@ class Word(unicode):
     translator = Translator()

-    def __new__(cls, string, pos_tag=None):
+    def get_instance(cls, string, pos_tag=None):
\end{lstlisting}

We use the following prompt:

\begin{mdframed}[backgroundcolor=gray!8, linecolor=black, linewidth=0.5pt]
You are a coding assistant specializing in refactoring. \\
You will receive a unified diff (git-style) that transforms an initial program into a resulting program. Use ONLY the information in the diff and general language semantics to judge one condition: \\
1) BEHAVIOR: The initial and resulting programs have the same observable behavior (same outputs, exceptions, externally visible state, timing/synchronization). \\
\\
ASSUMPTIONS (apply strictly): \\
- Code not shown in the diff is unchanged. \\
- If a symbol is renamed and all references *touched by the diff* are updated consistently, assume other call sites (not shown) are also updated unless the diff provides concrete evidence to the contrary (e.g., moved/renamed without updating a same-file reference). \\
- Ignore comment-only and whitespace-only changes. \\
- Treat added logging/printing, synchronization, exception type changes, constant value changes, and evaluation-order changes as potentially behavior-changing. \\
- Do not invent missing files or symbols; base conclusions on what the diff shows plus language rules (e.g., method overriding rules, access modifiers, package/module visibility, generics bounds, checked exceptions). \\
\\
ANALYSIS STEPS: \\
A. Parse the diff to extract concrete edits per file: renames/moves, visibility changes, signature changes (names, params, generics, exceptions), body edits, added/removed fields/methods/annotations, imports/package/module edits, synchronization/volatile/final changes, constant literal changes, control-flow edits, and I/O/logging. \\
B. BEHAVIOR checks (evidence-based): \\
- Control-flow or evaluation-order changes (e.g., moving side-effecting expressions into arguments), side effects, mutable state writes, synchronization/locking changes, exception \\  type/message/throwing-site changes, return value or constant changes, equals/hashCode/compareTo/serialization changes, annotation retention/values that affect runtime, static initialization reordering, I/O/logging added/removed if the API contract includes them. \\
- Pure renames/moves with consistent updates and no semantic edits to method bodies are behavior-preserving. \\
\\
OUTPUT FORMAT (must be exact): \\
- If the condition holds, respond with EXACTLY: \\
YES \\
- If it fails, respond with EXACTLY: \\
NO \\
After the first line, provide a brief explanation (1–3 sentences) describing the main evidence for failure. If you answered YES, provide no further text. \\
Answer in plain text. No JSON, no tool calls. \\
\\
Analyze this unified diff and answer strictly following the OUTPUT FORMAT above. \\
\code{diff}
\end{mdframed}

\subsection{Categorizing Behavioral Changes and Bugs}
\label{subsec:techStep7}

\paragraph{Categorizer of Behavioral Changes.} The categorizer stores, for each analyzed instance, the refactoring type, the git-style diff, the model, the prompt, the model response, and the resulting metrics. It then checks the response for predefined keywords that indicate whether behavioral preservation holds and emits the corresponding classification. This structured output supports aggregation, filtering, and comparison across instances. The consolidation of repeated failing instances into distinct bug reports is performed afterward.

\paragraph{Bug Reports.} In the last step, we inspect the instances classified as behavior-changing and group together those that share the same refactoring type, failure symptom, and minimal reproducer. Each group becomes one distinct bug candidate. To make bug reports easier to inspect, we reduce large programs using a delta-debugging-inspired process~\cite{pontes-fse-2019,Zeller-book-2009}: we iteratively remove code, reapply the refactoring, and keep only the constructs needed to reproduce the same behavioral change. For example, we reduced code from the TextBlob repository\footnote{\url{https://bit.ly/textBlobFile}} to obtain the motivating example from Section~\ref{sec:motivatingExample}.

\section{Evaluation}
\label{sec:evaluation}

This section presents the research questions, methodology, results, and discussion of our behavioral change detection evaluation.

\subsection{Definition}
\label{subsec:rqsBC}

Following the Goal-Question-Metric template~\cite{Basili1994}, we evaluate our oracle with respect to its ability to detect behavioral changes in Python refactoring implementations. We address the following research questions:
\begin{description}
    \item[\textbf{RQ$_{1}$}] To what extent can the oracle detect behavioral changes in Python refactoring implementations?
    To answer this question, we count the number of transformations applied by refactoring implementations that introduce behavioral changes.
    \item[\textbf{RQ$_{2}$}] What categories of behavioral changes are exposed by our technique?
    To answer this question, we classify the distinct behavioral changes found by our technique.
    \item[\textbf{RQ$_{3}$}] To what extent are the reported bugs accepted by developers?
    To answer this question, we inspect the issue-tracker evidence associated with the reported bug candidates, including maintainer comments, label changes, and issue resolution status, and count how many reports were accepted as bugs.
\end{description}
Unless otherwise stated, the answers to RQ$_1$--RQ$_3$ are based on the main evaluation conducted with \oss{}.

\subsection{Methodology}
\label{subsec:planningBC}

\subsubsection{Subject}

We selected the \emph{TextBlob} project, version~0.17.1, as the subject of our evaluation. TextBlob is a natural-language processing library for Python (compatible with versions 2 and 3), consisting of over 3,000 lines of code. It has also been used in prior studies~\cite{dilhara-ICSE-2022,Tsantalis-ICSE-2018}. We chose TextBlob because it is a real-world Python project with sufficient syntactic diversity to exercise multiple refactoring scenarios while still remaining small enough to support the required manual inspection and bug reduction steps.
To evaluate the oracle, we used a human-expert baseline as ground truth. A Python developer with ten years of professional experience acted as the sole judge and received the same input provided to the model: a git-style unified diff together with the behavioral-preservation criteria from Section~\ref{subsec:techStep6}. The expert saw only the diff, labeled each transformation as behavior-preserving or behavior-changing, and recorded a brief justification. We then compared the oracle's predictions against these labels and report accuracy, precision, recall, and F1-score, treating any disagreement as an error.

\subsubsection{Refactoring Implementations}

We evaluated Rope 1.3.0~\cite{rope} refactoring types that are common in practice~\cite{Murphy-Hill-TSE-2012,golubev-fse-2021}: Rename Field, Rename Method, Inline Method, Extract Method, Move Field/Method, and Use Function. These transformations expose Python-specific risks such as reflection, dynamic dispatch, operator overloading (e.g., \code{\_\_lt\_\_}, \code{\_\_add\_\_}), loss of object context (\code{self}), and name shadowing. Extending the evaluation to additional refactoring types mainly requires implementing the corresponding Step~1 automation.

\subsubsection{Tooling, Environment and Procedure}

All experiments were performed locally using Ollama~\cite{langchain-ollama-2025} on a PC with 64 GB RAM and an NVIDIA RTX 4060 GPU with 8 GB VRAM.
We used \oss{}, developed by OpenAI~\cite{openai-gptoss-2025}.
Because the model does not fully fit in VRAM, Ollama used hybrid CPU--GPU execution.

\revision{We reused 217 program pairs produced with Rope 1.3.0 in our prior study~\cite{sbes2025}.}
\revision{In that study, TextBlob modules were parsed with Python's AST, candidate methods and fields were selected for each refactoring type, and Rope generated the refactored programs.}
\revision{Here, we retained pairs with successful transformations and no reported issues, generated a git-style diff for each, submitted each diff to the foundation-model oracle and human-expert baseline, and compared their labels using classification and stability metrics.}
\revision{For transformations labeled behavior-changing, we manually grouped recurring failures by root cause, reduced representative examples, and reported the resulting bug candidates. We used this real-project protocol instead of a manually curated benchmark to study refactoring implementations on naturally occurring Python code under realistic transformation conditions.}

\subsubsection{Metrics}

To evaluate the technique's performance, we employ standard binary-classification metrics: precision, recall, F1-score, and accuracy, treating the presence of a behavioral change as the positive class. In our context, recall is particularly important because missing a behavioral change may leave a refactoring bug undetected, whereas precision helps limit false alarms.

To assess the reliability of foundation models across repeated attempts, we additionally adopt the accuracy and stability metrics presented in Table~\ref{tab:metrics}. These metrics capture not only whether a model produces correct results, but also whether it does so consistently across multiple executions, an essential property for tools that developers must trust.

\begin{table}[t]
\footnotesize
\caption{\mbox{Accuracy and stability metrics used in our study~\protect\cite{atil2025nondeterminismdeterministicllmsettings,chen-openai-2021}.}}
\begin{tabularx}{\columnwidth}{ p{2.25cm} X}
\toprule
\textbf{Metric} & \textbf{Definition} \\
\midrule
Mean Accuracy & Rate of correct answers across \emph{k} responses for one question, averaged over all questions. \\
Accuracy Spread & Difference between the maximum and minimum mean accuracy across \emph{k} attempts. Lower values indicate more stable accuracy. \\
\emph{pass@\(k\)} & Returns 1 if at least one of the \emph{k} responses is correct; 0 otherwise. Averaged over all questions. \\
\emph{tar@\(k\)} & Returns 1 if all \emph{k} responses are identical, regardless of correctness; 0 otherwise. Averaged over all questions. \\
\emph{cons@\(k\)} & Returns 1 if the most frequent response across \emph{k} attempts is correct; 0 otherwise. Averaged over all questions. \\
\bottomrule
\end{tabularx}
\label{tab:metrics}
\end{table}

Among these measures, mean accuracy and \emph{pass@\(k\)} characterize how often the oracle reaches the correct decision across repeated attempts, while accuracy spread, \emph{tar@\(k\)}, and \emph{cons@\(k\)} characterize how stable those answers remain from run to run. Together, they provide a compact view of both correctness and self-consistency, which is especially useful when evaluating an oracle intended to support developers in practice.

\subsubsection{Prompt and Responses}

We evaluated \oss{} with zero-shot prompting~\cite{prompts,prompt-techniques,zero-shot-prompt} and $k=3$ repeated runs, using the default Ollama API configuration~\cite{langchain-ollama-2025}. Each query consisted \revision{of the prompt described in Section~\ref{subsec:techStep6} together with} the git-style diff \revision{of one transformation instance.} Listing~\ref{lst:motivate-diff} \revision{illustrates the diff format}. We manually analyzed all responses.
Classification metrics are computed from the first-line YES/NO labels. Explanation correctness is assessed separately.

\subsection{Results}
\label{subsec:resultsBC}

\subsubsection{RQ$_1$. Detecting behavioral changes}

\revision{For each input, we executed the approach three times and used the modal response, that is, the most frequently produced prediction, as the final output for the evaluation. Based on these aggregated predictions, our approach achieved a precision of 0.67, a recall of 0.91, an accuracy of 0.84, and an F1-score of 0.77. These results indicate that the oracle detects behavioral changes in real-world Python refactoring instances with high recall and good overall accuracy, although it still produces some false positives.}
Listing~\ref{lst:rq1:fp} presents a diff for the application of the Inline Method refactoring.
According to the Inline Method definition~\cite{Fowler-book-1999}, the refactoring must replace each call with the function's body and remove the function definition.
In this case, the diff does not show any call replaced with the method body. This indicates that the function is not used in other code locations shown in the diff.

Our oracle identified the function removal but explored scenarios not present in the diff, as shown next.

\begin{tcolorbox}[
    colback=gray!8,
    colframe=black!70,
    fonttitle=\bfseries,
    sharp corners,
    boxrule=0.6pt,
    top=0.2mm,
    bottom=0.2mm,
    before skip=1pt,
    after skip=1pt
]
\noindent
The diff removes `\code{upper()}' method from `\code{WordList}', altering the public API. Code that previously called `\code{WordList. upper()}' would now either fail at runtime or produce a different result, changing the program’s observable behavior.
\end{tcolorbox}

\begin{lstlisting}[
  language=diff,
  caption={A git-style diff after applying the Inline Method refactoring.},
  label={lst:rq1:fp},
  belowskip=-5pt
]
--- a/blob.py
+++ b/blob.py
@@ -289,10 +289,6 @@ class WordList(list):
             self.append(e)

-    def upper(self):
-        """Return a new WordList with each word upper-cased."""
-        return self.__class__([word.upper() for word in self])
-
     def lower(self):
\end{lstlisting}

In this case, our technique yields a false positive under the diff-scoped baseline. The use of an external API is not evident in the diff, so the expert classifies the transformation as behavior-preserving when considering only the edited code shown.
As another example, Listing~\ref{lst:rq1:bug} shows an application of the Rename Method in the function \code{\_\_new\_\_}.
Rope allowed the refactoring to be applied to a special method that customizes the instantiation logic of a class.

\begin{lstlisting}[
  language=diff,
  caption={A git-style diff after applying the Rename Method refactoring.},
  label={lst:rq1:bug},
  belowskip=-6pt
]
--- a/blob.py
+++ b/blob.py
@@ -67,7 +67,7 @@ class Word(unicode):
     translator = Translator()
-    def __new__(cls, string, pos_tag=None):
+    def tags(cls, string, pos_tag=None):
\end{lstlisting}

Our approach identified the applied change, highlighting the risk of renaming special methods, as shown next.

\begin{tcolorbox}[
    colback=gray!8,
    colframe=black!70,
    fonttitle=\bfseries,
    sharp corners,
    boxrule=0.6pt,
    top=0.2mm,
    bottom=0.2mm,
    before skip=1pt,
    after skip=1pt
]
\noindent
The diff renames the special constructor method `\code{__new__}' to `\code{tags}', removing the custom instantiation logic that handled the `\code{pos_tag}' argument. As a result, constructing a `Word' instance will now use the default `\code{object.__new__}', changing object creation semantics and likely altering program behavior.
\end{tcolorbox}

As a complementary case, Listing~\ref{lst:rq1:fn} presents an application of the Extract Method refactoring.
This scenario exposes a failure related to Python’s indentation-based scoping rules.
Unlike other programming languages that use braces or keywords, Python uses indentation to determine which statements belong together.
In this sample, the oracle also struggled with this case and incorrectly predicted behavior preservation.

\begin{lstlisting}[
  language=diff,
  caption={A git-style diff after applying the Extract Method refactoring.},
  label={lst:rq1:fn},
  belowskip=-6pt
]
--- a/formats.py
+++ b/formats.py
@@ -95,7 +95,10 @@ class CSV(DelimitedFormat):
 class TSV(DelimitedFormat):
-    delimiter = "\t"
+    delimiter = extracted_method()
+def extracted_method():
+    return "\t"

 class JSON(BaseFormat):
\end{lstlisting}

\subsubsection{RQ$_2$. Identifying behavioral change bugs}
To answer RQ$_2$, we characterize the categories of behavioral changes exposed by our approach.
\revision{Among the 217 analyzed transformation pairs, our expert baseline identified 64 behavioral-change instances, corresponding to refactoring cases in which the transformed program was judged to exhibit a behavioral change.}
\revision{These behavioral-change instances were distributed as follows: 52 Inline Method instances, 2 Rename Method instances, 1 Move Field instance, 7 Extract Method instances, 2 Rename Field instances, and no instance for Move Method and Use Function.}
We consolidated \revision{them} by grouping instances that shared the same refactoring type, the same primary symptom, and the same reduced reproducer, resulting in \totalBugsBehavioral{} distinct behavioral changes. These behavioral changes fall into recurring categories tied to Python-specific mechanisms:
(i) \emph{Runtime failures due to incorrect name/import handling} (e.g., introducing \texttt{NameError} or referencing non-existent imports); 
(ii) \emph{Incorrect transformation of expressions/call sites} (e.g., incorrect parameter name when rewriting string formatting); 
(iii) \emph{Violations of Python data model constraints} (e.g., applying transformations to special ``dunder'' methods or rich comparison methods that frameworks implicitly rely on); 
(iv) \emph{Inheritance/contract-related violations} (e.g., removing overridden or abstract methods whose presence affects runtime behavior); 
and (v) \emph{Structural dependency breakage} (e.g., introducing circular dependencies). 
Therefore, RQ$_2$ shows that the detected bugs are not isolated anomalies, but recurring classes of Python-specific behavioral failures.

\subsubsection{\textbf{RQ$_3$.} Developer acceptance of reported bugs}
The \totalBugsBehavioral{} behavioral changes from RQ$_2$ were reported to Rope as GitHub issues; Table~\ref{table:rq5} summarizes the reports. They span five refactoring types: one for Extract Method, seven for Inline Method, one for Move Field, two for Rename Field, and two for Rename Method. Issues 4, 7, and 8 had also been observed in a separate manual analysis, whereas the remaining ten issues were uncovered exclusively by the behavioral-change detection technique. To assess acceptance, we inspected maintainer comments, label changes, and issue status. We considered a report accepted when it remained classified as a bug and no maintainer feedback rejected it as a defect; otherwise, we treated it as not accepted.
According to this criterion, 12 of the 13 reported bugs (92.3\%) were accepted by Rope developers. The only non-accepted case was issue \#750 (Table~\ref{table:rq5}, ID 5), for which a maintainer removed the bug label, reclassified the report as an enhancement, and stated that inlining \code{\_\_new\_\_} is currently unsupported rather than buggy. Therefore, RQ$_3$ indicates that most of the reported bug candidates were accepted by developers under our issue-tracker-based criterion.

\begin{table}[t]
\centering
\scriptsize
\setlength{\tabcolsep}{2pt}
\caption{Summary of the behavioral-change bugs reported to Rope.}
\begin{tabularx}{\columnwidth}{@{}r >{\RaggedRight\arraybackslash}p{1.41cm} X r@{}}
\toprule
\textbf{ID} & \textbf{Refactoring} & \textbf{Issue} & \textbf{Tracker} \\
\midrule
1  & Extract Method & Extracted method defined outside class scope, causing `NameError' & 825 \\
2  & Inline Method  & Removed test-file import, causing a runtime error & 826 \\
3  & Inline Method  & Failed to inline all attributes, causing `NameError' & 827 \\
4  & Inline Method  & Incorrect parameter name when rewriting the formatted string & 757 \\
5  & Inline Method  & Allows inlining datamodel methods, breaking collection/framework behavior & 750 \\
6  & Inline Method  & Removes an overridden method, changing the stored-object type & 828 \\
7  & Inline Method  & Removes a rich comparison method & 758 \\
8  & Inline Method  & Allows inlining abstract methods & 744 \\
9  & Move Field     & Introduces a circular dependency & 829 \\
10 & Rename Field   & Allows object datamodel names & 830 \\
11 & Rename Field   & Rewrites an import and adds a nonexistent name & 831 \\
12 & Rename Method  & Allows renaming to datamodel function names & 832 \\
13 & Rename Method  & Allows renaming functions to datamodel function names & 833 \\
\bottomrule
\end{tabularx}
\label{table:rq5}
\end{table}

\subsection{Discussion}
\label{subsec:discussionBC}

\subsubsection{Test Input Programs}

For our evaluation, we selected the open-source project TextBlob, which has also been used in prior studies~\cite{dilhara-ICSE-2022,Tsantalis-ICSE-2018}. TextBlob includes 77\% of Python's keywords, which helped expose multiple refactoring bugs, but it does not cover \code{async}, \code{await}, \code{del}, \code{with}, \code{nonlocal}, \code{global}, \code{finally}, and \code{yield}. Consequently, our study may miss behavioral changes tied to those constructs.
For example, an Extract Method that moves an update to a global variable into a helper without preserving the \code{global} declaration would change the program output from \code{True} to \code{False}; our technique would flag such a case if it appeared in the subject program.

\subsubsection{Bugs}

We assessed bug acceptance using maintainer comments, label changes after triage, and issue resolution status, rather than relying on labels added by the reporter at issue creation time. Subsequent maintainer feedback is our primary evidence for whether a report was considered a valid bug, reclassified, or left open.
The reporting process also helped improve later submissions by clarifying reproduction steps and expected versus observed behavior. Based on Table~\ref{table:rq5}, we recommend adding preconditions for special dunder methods and abstract methods, validating name and import resolution after transformations, and checking circular dependencies in Move refactorings.

\subsubsection{False Positives}

We denote as a false positive any case in which our oracle predicts a behavioral change while the expert identifies the transformation as behavior-preserving. Conversely, a false negative occurs when the oracle predicts behavior preservation but the expert identifies a behavioral change. More generally, any mismatch between the oracle and the expert constitutes a disagreement.
A recurring source of such discrepancies is the way Python uses indentation to define the scope of classes and functions. This syntactic characteristic made it particularly difficult for our oracle, which reasons over git-style diffs, to interpret certain transformations correctly. In Python, the scoping of a declaration depends on the exact number and type of whitespace characters (spaces or tabs) preceding it. For the underlying base model, this requires correctly parsing and semantically interpreting whitespace in the diff, rather than relying only on more salient lexical tokens. In contrast, a human expert can quickly infer scope by visually inspecting the indentation structure of the code.

\subsubsection{Other Tools}

We conducted an exploratory manual check to determine whether the bugs detected in Rope also manifested in popular IDEs such as PyCharm Community 2025.2.4, PyDev 10.1.4, and VSCode 1.106.2. In this verification, we considered only those bug reports whose refactoring type was actually supported by each IDE. For each eligible case, we reused the simplified input code from the bug report together with the same parameters outlined in the original report, manually applied the corresponding refactoring in the IDE, and then analyzed the resulting transformation. No analogous problems were observed in these IDE environments.
Nevertheless, our technique is designed to be portable to new settings, requiring changes only in the automation of refactoring instance application in Step 1 of our pipeline. The subsequent steps remain unchanged across tools, which demonstrates both the adaptability of our approach and its potential for integration with a wide range of refactoring tools and development environments.

\subsubsection{Results over multiple runs}

We executed our approach with \oss{} for $k \in \{1, 2, 3\}$ repeated runs on the 217 refactoring instances~\cite{leviathan2025promptrepetitionimprovesnonreasoning}. Running all three passes required nearly 9 hours.
\revision{Table~\ref{fig:metrics} shows that across $k \in \{1,2,3\}$ repeated runs the oracle achieved a mean accuracy of 0.799.} As additional attempts are allowed, \emph{pass@k} increases to 0.871 for $k = 2$ and 0.903 for $k = 3$, showing that repeated querying raises the chance of obtaining a correct decision.
Accuracy remained stable across runs, ranging from 0.793 to 0.810, with an accuracy spread of at most 0.02. The \emph{tar@k} values decrease from 1.0 at $k = 1$ to 0.779 at $k = 3$, reflecting more varied answers. \revision{For $k=3$, \emph{cons@k} reaches 0.839, which corresponds to the modal-response accuracy reported in RQ$_1$.}
Repeated runs improve effectiveness while maintaining acceptable stability.

\begin{table}[tb]
\footnotesize
\centering
\caption{Accuracy and stability metrics for \oss{} across $k$ repeated runs on the 217 refactoring instances.}
\begin{tabular}{lccc}
\toprule
\textbf{Metric/k} & \textbf{1} & \textbf{2} & \textbf{3} \\
\midrule
\emph{tar@k} & 1.000 & 0.839 & 0.779 \\
\emph{cons@k} & 0.793 & 0.719 & 0.839 \\
\emph{pass@k} & 0.793 & 0.871 & 0.903 \\
Accuracy & 0.793 & 0.795 & 0.810 \\
\bottomrule
\end{tabular}
\label{fig:metrics}
\end{table}

\subsubsection{GPT-5.2 Thinking}
\label{subsec:gpt52thinking}

\paragraph{Setup.}
To explore how a state-of-the-art proprietary model behaves on the same task, we additionally evaluated \textsc{GPT-5.2 Thinking}, which ranks among the top proprietary models according to the LLM Arena benchmark~\cite{chiang2024chatbot}. This analysis is complementary rather than a direct baseline replacement, because the proprietary model exhibited a broader reasoning scope than the strictly diff-scoped evaluation protocol adopted for \oss{}. Accordingly, it is separate from RQ$_1$--RQ$_3$, which are based on \oss{}.

To control evaluation costs, we conducted this analysis through the web-based chat interface of \textsc{GPT-5.2 Thinking} using a personal account with a free offer to access the \textsc{Plus} plan. Executing the model required manually pasting the 217 prompts and collecting the answers, a process that took approximately 72 hours due to page refreshes, prompt submission, and response collection latency. We applied the same prompts used with \oss{} (Section~\ref{subsec:techStep6}) and set $k = 1$ to maintain feasibility under these constraints. We adapted some symbols and removed blank lines to comply with the chat interface's formatting limitations, but the semantic content of each prompt remained unchanged.

\paragraph{Overall Outcome.}
Out of the 217 refactoring instances, \textsc{GPT-5.2 Thinking} classified 197 (90.8\%) as behavioral changes (NO) and 20 (9.2\%) as behavior-preserving (YES). All 20 YES instances agreed with the expert baseline, and every instance labeled as a behavioral change by the expert was also flagged by the model. However, 133 of the 197 NO responses correspond to cases where the expert labeled the transformation as behavior-preserving. Rather than reporting standard classification metrics, which would be misleading given the scope mismatch between the model's analysis and the diff-based expert baseline, we conduct a qualitative analysis of these disagreements.

\paragraph{Response Format.}
The model adhered strictly to the requested format. All 20 YES responses consisted only of the word ``YES''. All 197 NO responses began with ``NO'' followed by a brief explanation, typically between one and three sentences (mean length of approximately 36 words). Approximately 45\% of the NO responses used definitive language such as ``will now fail'' or ``will raise an \code{AttributeError}'', while approximately 35\% used conditional formulations such as ``can change behavior'' or ``can raise an \code{ImportError}''.

\begin{lstlisting}[
  language=diff,
  caption={A git-style diff by applying Extract Method on \code{\_\_repr\_\_}.},
  label={lst:gpt52:conditional:subclass},
  belowskip=-1pt,
]
--- a/mixins.py
+++ b/mixins.py
@@ -58,12 +58,15 @@ class StringlikeMixin(object):
     def __repr__(self):
-        class_name = self.__class__.__name__
+        class_name = self.extracted_method()
         text = self.__unicode__().encode("utf-8") if PY2 else str(self)
         ret = '{cls}("{text}")'.format(cls=class_name, text=text)
         return binary_type(ret) if PY2 else ret

+    def extracted_method(self):
+        return self.__class__.__name__
\end{lstlisting}

\paragraph{\revision{Reasoning Categories.}}
A manual review of the GPT-5.2 responses compared each model answer with the diff-scoped expert baseline.
Out of the 217 refactoring instances, GPT-5.2 produced 197 NO responses and 20 YES responses.
The 197 NO responses comprise 64 cases that agree with the expert baseline and 133 cases in which GPT-5.2 flagged a behavioral change but the expert classified the transformation as behavior-preserving under the diff-scoped protocol.
We therefore distinguish baseline-confirmed behavioral changes from project-aware risk claims that go beyond the baseline.

\textbf{(1) Behavior preserved} (20 instances).
These correspond to the YES responses, in which the model determined that the transformation does not alter the program's observable behavior.
All 20 cases agreed with the expert baseline.

\textbf{(2) Baseline-confirmed behavioral changes} (64 instances).
In these cases, GPT-5.2 answered NO and agreed with the diff-scoped expert baseline.
These responses correspond to transformations whose behavioral impact was visible under the same diff-scoped protocol used in the main evaluation.
The main recurring patterns include runtime failures caused by name or import handling, import-time or definition-time errors, removal or modification of methods used in the transformed code, violations of Python data-model behavior, and changes in logic, return values, or observable output.

Listing~\ref{lst:rq1:fn} (Section~\ref{subsec:resultsBC}) illustrates a case in this category.
The Extract Method refactoring places the extracted function after the class definition that depends on it, causing a \code{NameError} at import time.
While \oss{} incorrectly classified this case as behavior-preserving, \textsc{GPT-5.2 Thinking} correctly identified the problem:

\begin{tcolorbox}[
    colback=gray!8,
    colframe=black!70,
    fonttitle=\bfseries,
    sharp corners,
    boxrule=0.6pt,
    top=0.2mm,
    bottom=0.2mm,
    before skip=1pt,
    after skip=1pt
]
\noindent
The class attribute \code{TSV.delimiter} is now set by calling \code{extracted\_method()} during class definition, but \\ \code{extracted\_method} is defined only afterward, causing a \\ \code{NameError} when the module is imported/executed.
\end{tcolorbox}

\textbf{(3) Baseline-disagreeing project-aware risk claims} (133 instances).
In these cases, GPT-5.2 answered NO, but the expert baseline classified the transformation as behavior-preserving under the diff-scoped protocol.
We do not count these cases as confirmed behavioral changes.
Instead, they represent project-aware risk claims: the model often reasons about effects that require information beyond the local diff, such as external callers, public API usage, subclass overrides, dynamic dispatch, introspection, wildcard imports, monkeypatching, or project-wide import interactions.

The risk categories in this group are overlapping rather than mutually exclusive.
They include risks from external access or introspection, risks from subclass overrides or dynamic dispatch, public API removal or modification, public API addition, and import-time effects.
Some responses are explicitly conditional, using formulations such as ``can change behavior'' or ``could raise an exception''.
Others are phrased as definitive claims, but still depend on assumptions about external clients or broader project context that were outside the expert baseline.

Listing~\ref{lst:gpt52:conditional:subclass} presents an example of the Extract Method refactoring where the model identifies a project-aware risk due to dynamic dispatch.
For this case, the model produced the following response.

\begin{tcolorbox}[
    colback=gray!8,
    colframe=black!70,
    fonttitle=\bfseries,
    sharp corners,
    boxrule=0.6pt,
    top=0.2mm,
    bottom=0.2mm,
    before skip=1pt,
    after skip=1pt
]
\noindent
\code{\_\_repr\_\_} now calls \code{self.extracted\_method()} instead of directly using \code{self.\_\_class\_\_.\_\_name\_\_}. As \\ \code{extracted\_method} is a normal instance method, subclasses (or other bases in the MRO) can override it, changing the class name used in \code{\_\_repr\_\_} and thus observable output.
\end{tcolorbox}

In this scenario, the model's reasoning is technically sound: introducing a new method that is invoked via dynamic dispatch opens the possibility that a subclass overrides \code{extracted\_method}, thereby altering the return value of \code{\_\_repr\_\_}.
However, the expert baseline, which evaluates behavioral equivalence based solely on the diff, classifies this transformation as behavior-preserving because no such subclass override is evident in the diff itself.
This example illustrates a limitation of the diff-scoped baseline: by not considering the broader project context, it cannot account for behavioral risks that arise from language-level mechanisms such as dynamic dispatch, even when these risks are well-documented sources of refactoring bugs~\cite{Soares-TSE-2013,Mongiovi-TSE-2018,Mongiovi-SBPL-2014,Oliveira-2023-IST,Soares-JSS-2013,Soares-IEEE-2010}.

Another recurring pattern in this group involves changes to the module's externally visible namespace.
For instance, when an Extract Method refactoring introduces a new top-level function, the model flags it as a behavioral risk:

\begin{tcolorbox}[
    colback=gray!8,
    colframe=black!70,
    fonttitle=\bfseries,
    sharp corners,
    boxrule=0.6pt,
    top=0.2mm,
    bottom=0.2mm,
    before skip=1pt,
    after skip=1pt
]
\noindent
The module now defines a public symbol \code{extracted} \code{\_method}, which changes the module's externally visible state (e.g., \code{dir()}, \code{from compat import *}, attribute access).
\end{tcolorbox}

This reasoning considers effects that are observable only when external code interacts with the modified module, a perspective that the diff-based baseline does not capture.
Steimann and Thies~\cite{Steimann-ecoop-2009} have shown that even mainstream Java IDEs fail to preserve accessibility during refactorings, leading to unintended behavioral changes, which suggests that the model's concerns about externally visible namespace changes are well-founded.

As a further example, Listing~\ref{lst:rq1:fp} (presented in Section~\ref{subsec:resultsBC}) shows the Inline Method refactoring removing the \code{upper()} method from \code{WordList}.
For this case, the model produced the following response.
The expert baseline classifies this instance as behavior-preserving, since the diff shows no remaining call sites for the removed method.
However, the model's reasoning considers that external callers of the public API could depend on this method.

\begin{tcolorbox}[
    colback=gray!8,
    colframe=black!70,
    fonttitle=\bfseries,
    sharp corners,
    boxrule=0.6pt,
    top=0.2mm,
    bottom=0.2mm,
    before skip=1pt,
    after skip=1pt
]
\noindent
The \code{WordList.upper()} method was removed. Any code that previously called \code{upper()} will now raise the error \code{AttributeError} instead of returning an upper-cased \code{WordList}, changing observable behavior.
\end{tcolorbox}

\paragraph{Scope Mismatch with the Baseline.}
The high number of disagreements between the model and the expert (133 out of 197 NO responses) reflects a mismatch between the scope of the model's analysis and the scope of the expert evaluation.
Our baseline was constructed by having the expert assess behavioral equivalence based solely on the information contained in the diff, consistent with the evaluation protocol used for \oss{}.
However, \textsc{GPT-5.2 Thinking} systematically reasons about effects that extend beyond the diff, such as consequences for external consumers of public APIs, potential subclass overrides via dynamic dispatch, introspection-visible namespace changes, and import-time effects in dependent modules.
These concerns correspond to categories of defects that have been confirmed as real bugs in refactoring engines by prior studies~\cite{Soares-TSE-2013,Mongiovi-TSE-2018,Wang-TOSEM-2025}, but they require a project-aware baseline before they can be counted as confirmed behavioral changes in this study.

Revising the baseline to consider the full project context could reclassify a portion of these disagreements, since many of the model's concerns, such as removed public methods or newly introduced namespace symbols, represent genuine risks in a real-world codebase.
Such a revision could also surface additional bug candidates that are currently masked by the conservative scope of the diff-based evaluation.
However, constructing a project-aware baseline requires the expert to analyze the complete codebase for each transformation, which significantly increases the effort and complexity of the evaluation process.
We leave this as future work.

The case presented in Listing~\ref{lst:rq1:fn}, incorrectly classified as behavior-preserving by \oss{}, was correctly identified by \textsc{GPT-5.2 Thinking}, which produced an accurate explanation of the root cause.
We did not find any scenario where \oss{} detects a behavioral change that \textsc{GPT-5.2 Thinking} classifies as behavior-preserving.
This observation, combined with the fact that no behavioral change went undetected by \textsc{GPT-5.2 Thinking}, suggests that the proprietary model adopts a more conservative and comprehensive  strategy, flagging a broader set of potential issues at the cost of producing more disagreements under a diff-scoped baseline.

\section{Threats to Validity}
\label{sec:threats}

Our evaluation involves some threats to validity. First, the behavioral-equivalence baseline is derived from the manual judgment of a single expert, and the subsequent inspection of oracle answers and consolidation of failing executions into distinct bug reports also required manual judgment. These steps may introduce subjectivity or occasional misclassification, especially in borderline cases. We mitigated this threat by requiring written justifications for the expert labels, by reducing each reported issue to a minimal reproducer, and by submitting the resulting bug reports to the tool maintainers for external inspection. Nevertheless, maintainer feedback \revision{only} validates the reported bug candidates rather than the full set of expert labels. \revision{Because precision, recall, accuracy, and F1-score are computed against this expert baseline, any ambiguity or error in the baseline can directly affect the reported metrics. Future replications should therefore include multiple independent evaluators, adjudication of disagreements, and inter-rater agreement measures.}

Second, our study is based on a single real-world Python project, TextBlob, and on a subset of refactoring types supported by Rope. Although TextBlob offers substantial syntactic diversity and these refactorings are widely used in practice~\cite{Murphy-Hill-TSE-2012}, the results may not generalize to all Python codebases, language features, or refactoring operations. Moreover, the 217 analyzed transformation pairs support a focused bug-finding and oracle-evaluation study, but they do not support broad prevalence claims about Python refactoring bugs. \revision{The concentration of behavioral-change instances in Inline Method also means that the aggregate metrics may reflect the characteristics of this refactoring type more strongly than those of the other refactoring types.} Also, our manual verification in other IDEs was exploratory rather than systematic, so it should not be interpreted as a comprehensive comparative evaluation. Additional projects, tools, IDEs, and refactoring types is part of our future work.

Third, the oracle is intentionally diff-scoped: both the model and the human baseline reason only over the edited code shown in the diff, treating unseen code as unchanged. This choice improves feasibility on real-world projects, but it may miss project-wide effects that depend on external callers, subclass hierarchies, or import-time interactions outside the diff. Likewise, Python's indentation-sensitive syntax makes some transformations harder to interpret from diffs alone, which contributed to some false positives and false negatives. Thus, our results characterize diff-local behavioral-change detection rather than full project-aware semantic equivalence. \revision{A project-aware baseline, or a diff enriched with selected project context, could change the classification of borderline cases involving public APIs, subclasses, or imports.}

Finally, our construct and conclusion validity are limited by the use of a single foundation model in the main evaluation, by the fact that our method targets only behavioral changes rather than all categories of refactoring bugs, and by the limited dataset size per refactoring type. The results may therefore reflect characteristics of \oss{}, as well as sensitivity to class imbalance and dependence among observations. The complementary GPT-5.2 analysis is exploratory and should not be interpreted as a controlled model comparison. \revision{We also did not compare the oracle against TextBlob's test suite, static analyses, or bytecode-based checks, so our evaluation does not determine how many of the detected behavioral changes could be found by these alternatives.} Our technique should be viewed as complementary to test suites and other validation techniques, not as a complete replacement. Future work should investigate other models, broader bug categories, more projects/tools, and stronger statistical reporting, such as confidence intervals.

\section{Related Work}
\label{sec:relatedWork}

Research on refactoring established the concept, automated its basic forms, and exposed the difficulty of guaranteeing behavior preservation through preconditions~\cite{Opdyke-SOOPPA-1990,Opdyke-PHD-1992,Roberts-PHD-1999,Tokuda-ASE-2001,Schafer-PLPV-2009}.
Recently, AlOmar et al.~\cite{AlOmar-infsof-2021} summarized the landscape of behavior-preservation techniques and highlighted challenges. Several studies have tested refactoring engine correctness in statically typed languages, such as Java.
Daniel et al.~\cite{daniel-fse-07} and Gligoric et al.~\cite{Gligoric-ICSE-2010,Gligoric-ecoop-13} used generated programs to exercise refactorings. Other studies focused on concrete failure modes in refactoring engines: accessibility violations, behavioral changes, compilation errors, and overly strong preconditions~\cite{Steimann-ecoop-2009,Soares-TSE-2013,Soares-ICSM-2011,Mongiovi-TSE-2018,Mongiovi-icsme-2014}.
Wang et al.~\cite{wang-2024,Wang-TOSEM-2025} further showed that refactoring bugs remain widespread in mature tools and their root causes. These studies provide evidence that refactoring engines can still be faulty. In contrast, our work targets Python refactorings, applies transformations to real code rather than synthetic inputs, and evaluates behavior preservation through a diff-based oracle.
These differences in language, input generation, and oracle design are relevant, but not direct experimental baselines for our setting.

This distinction is important for Python.
Sch\"{a}fer~\cite{Schafer-fwrt-2012} emphasized the difficulty of validating refactorings in dynamic languages, where dynamic dispatch and other mechanisms make behavior preservation harder to guarantee. Recent work on Python refactoring shows that existing tools still face practical limitations.
Wang et al.~\cite{Wang-icsme-2025} report limitations in Python refactoring support, including restricted tool coverage and practitioner concerns on robustness.
Oliveira et al.~\cite{sbes2025} studied faulty Python refactorings and exposed problems related to type errors and other statically observable failures. Our work addresses a complementary problem: transformations that complete successfully and show no prior problem indication may still introduce behavioral changes. Rather than detecting explicit failures or type problems, we analyze apparently successful transformations and ask whether they are behavior-preserving edits.

The paper also relates to studies on refactoring practice and tool adoption.
Large empirical studies have shown that developers often avoid automated refactoring when tool support is inadequate, and that even in industrial settings refactorings are frequently performed manually~\cite{Tempero-ACM-2017,Kim-TSE-2014}.
More recently, Horikawa et al.~\cite{horikawa2025agenticrefactoringempiricalstudy} showed that refactoring is also common in code changes produced by AI coding agents.
Our work complements these studies by focusing on an orthogonal problem: assessing whether automated refactoring transformations preserve behavior.

LLMs have recently been used both to perform refactorings and to support refactoring-related analysis.
Midolo et al.~\cite{midolo2026humanmachinerefactoringassessing} evaluated LLMs as Python refactoring agents and relied on unit tests to assess preservation.
Dong et al.~\cite{dong-icse-2025} used LLMs to generate test programs for testing Java refactoring engines.
Gheyi et al.~\cite{gheyi2025evaluatingeffectivenesssmalllanguage} evaluated small language models as oracles for known refactoring bugs in Java and Python.
\revision{Gheyi et al.~\cite{gheyi2026foundationmodelsoraclesrefactoring} evaluated foundation models as oracles for detecting refactoring correctness issues in Java programs, using zero-shot prompting over real refactoring bugs from mature IDEs.}
These studies show that language models can support refactoring and refactoring validation, but they either focus on generating transformations, generating tests, or classifying known bug scenarios.
Our work differs by using a foundation-model oracle to analyze git-style diffs produced from real Python refactoring instances.
This enables a lightweight validation strategy that remains focused on the actual edits introduced by the transformation and does not require executing the program or generating tests.
In this sense, our approach is complementary to unit-test-based validation and to generated-program approaches.

Other recent studies use LLMs to suggest, refine, or validate code transformations in Python.
Examples include PyCraft~\cite{dilhara-fse-2024}, EM-Assist~\cite{Pomian-fse-2024}, LLM-guided refactoring recommendation approaches~\cite{Zhang-ins-2024,piao2025refactoringllmsbridginghuman}, and systems that simplify or restructure Python code with model assistance~\cite{DBLP:conf/apsec/ShirafujiOSMW23}.
These works focus primarily on producing or improving transformations, whereas our contribution is an oracle for checking whether a proposed transformation introduces a behavioral change.
This makes our technique a natural complement to systems that generate or apply refactorings.

\section{Conclusions}
\label{sec:conclusion}

We \revision{extended} \toolname{} \revision{with} a model-based approach for detecting behavioral changes introduced by Python refactorings.
By analyzing 217 transformations, \toolname{} identified \totalBugsBehavioral{} distinct bugs across the 7 refactorings studied.
These transformations are apparently successful, making the detected bugs quite relevant as they were not exposed by explicit tool failures or prior problem indications. Twelve of the 13 reported bugs reported bugs were accepted by developers.  
These results show that a diff-based foundation-model oracle can expose behavioral change failures in Python with good accuracy.
For developers, this reinforces the need to validate refactoring results rather than relying on full tool automation.
For tool builders, it highlights the need for stronger preconditions and validation mechanisms, especially in dynamic languages such as Python, which features challenging mechanisms such as dynamic dispatch, data-model methods, name resolution, and inheritance contracts.

Our results suggest that foundation models can be a valuable complement to traditional refactoring-validation techniques, where behavioral changes are difficult to express as static rules or detect with existing checks. Also, the observed variability, cost, and diff-scoped limitations indicate that these models are best used as an additional validation layer, rather than as a replacement for static and dynamic analysis techniques.
Thus, \toolname{} should be interpreted as a practical bug-finding aid for suspicious refactoring transformations, not as a complete correctness oracle. Our approach benefits from the compactness of git-style diffs, which makes large-model analysis feasible without requiring the full project as input, but it remains limited by model cost and by the scope of diff-based reasoning. Future work includes evaluating additional  refactoring tools, covering more refactoring types, and exploring multi-model or agent configurations for validation and repair.

\section*{Artifact Availability}
All study artifacts are available online~\cite{artifacts}.

\section*{Acknowledgments}
We want to thank the anonymous reviewers for their insightful suggestions. This work was partially supported by CNPq (306026/2026-0, 408040/2025-4, 403719/2024-0), CAPES (88887.313474/2026-00), FAPESQ-PB (268/2025), and INES.IA (National Institute of Science and Technology for Software Engineering Based on and for Artificial Intelligence), www.ines.org.br, CNPq grant 408817/2024-0.

\end{document}